# On an Approach to Bayesian Sample Sizing in Clinical Trials

by

## Robb J. Muirhead[1] and Adina I. Şoaita[2]

*We are honored to contribute to this Festschrift celebrating Professor Morris L. Eaton, one of the preeminent theoretical statisticians of our time. We pay tribute to him -- for his distinguished and illustrious career at the University of Minnesota, for his innovative and highly influential body of research, and for his dedicated and selfless service to the statistics profession.*

*Readers of this volume will be familiar with Joe Eaton's pioneering and fundamental research in many areas of theoretical statistics. They may be less familiar with his work on methodological problems of particular importance to statisticians in the pharmaceutical industry, and we would like to take this opportunity to highlight some of this work. He has made important contributions to our understanding of statistical issues in both early and late stage drug development; these will have lasting impact. They include research in very diverse areas (the following list is not intended to be exhaustive): dissolution profile testing, where the aim is to show through in vitro testing that the dissolution properties of (for example) a new formulation of an approved drug are statistically "similar" to the existing formulation, a necessary step in obtaining a bio-waiver for in vivo testing (Eaton et al, 2003); the statistical evaluation of a decision rule in a regulatory guidance for assessing whether a new drug in development prolongs the length of the QT interval (an ECG measurement) which, because QT prolongation is a safety signal for potential cardiac problems, can lead to the development of a compound being halted (Eaton et al, 2006a); the construction and evaluation of multivariate Bayesian predictive distributions and reference (or tolerance) regions for assessing a vector of laboratory values, providing a novel approach to monitoring the safety of a drug in clinical trials (Eaton et al, 2006b); the testing of multiple co-primary endpoints, where a new drug must demonstrate efficacy on a number of variables in order to gain regulatory approval (Eaton and Muirhead, 2007); and large-sample properties of "proper Bayesian" and "hybrid classical-Bayesian" approaches to sample sizing of clinical trials (Eaton et al, 2012, and used in the current paper). On behalf of the pharmaceutical statistical community, we thank Joe for his willingness to be involved in industry problems and for his many important contributions.*

---

[1] Statistical Consultant, Lyme CT
[2] Pfizer Inc, Groton CT

**Abstract**

This paper explores an approach to Bayesian sample size determination in clinical trials. The approach falls into the category of what is often called "proper Bayesian", in that it does not mix frequentist concepts with Bayesian ones. A criterion for a "successful trial" is defined in terms of a posterior probability, its probability is assessed using the marginal distribution of the data, and this probability forms the basis for choosing sample sizes. We illustrate with a standard problem in clinical trials, that of establishing superiority of a new drug over a control.

## 1. Introduction

Sample size determination plays an important role in the design aspect of studies in many fields, and especially in the planning of clinical trials. Although applicable to other situations (such as equivalence trials and non-inferiority trials), our focus here is on superiority trials, where the aim is to establish that a new experimental treatment is more efficacious than (superior to) a control, such as placebo or an existing treatment. The frequentist approach is well known. For a given test of the null hypothesis of no difference between the two treatments, we specify a significance level (or size) $\alpha$, the magnitude $\delta^*$ (say) of the magnitude of the treatment effect $\delta$ considered "clinically meaningful", and $1-\beta$, , the desired power at $\delta^*$. We then determine a sample size so that the test of level $\alpha$ will reject the null hypothesis with probability $1-\beta$ if, in fact, the "true" treatment effect is $\delta^*$. Frequentist sample size formulas are available in commonly occurring situations, whether based on standard tests themselves or on lengths of confidence intervals. (See, for example, Joseph and Bélisle (1997) and Whitehead et al (2008).) These formulas generally involve "guesses" (estimates) of the values of nuisance parameters, such as unknown variances. These estimates, which are usually obtained from prior studies, or from published studies on the same (or a similar) drug, are treated as "correct" in frequentist sample sizing (although, of course, sensitivity analyses are routinely carried out to see how sample sizes are affected by the choice of the values of nuisance parameters). The fact that the estimates obtained from prior studies and from expert opinions sometimes underestimate the variability actually observed in a new trial can lead to an underpowered study. In addition, what $\delta^*$ should actually represent remains open to debate. The regulatory guideline ICH E9 (which immediately follows introductory remarks by Lewis (1999)) recommends that $\delta^*$ may be based either on "a judgment concerning the minimal effect which has clinical relevance in the management of patients or on a judgment concerning the anticipated effect of the new treatment, where this is larger." That the anticipated effect size is often optimistically overstated may also lead to an underpowered study. For further discussion on these matters, see e.g. Grouin et al (2007) and O'Hagan et al (2005).

Because of the uncertainty involved in the specification of both $\delta^*$ and nuisance parameters, it is natural to incorporate Bayesian concepts into the sample size determination process. Possible approaches have been classified by Spielgelhalter et al (2004) as "hybrid classical-Bayesian", "decision-theoretic Bayesian", and "proper Bayesian". The most common example of the hybrid classical-Bayesian approach involves specifying a prior distribution for unknown model parameters, and then averaging the (frequentist) power function with respect to this prior to obtain an "average power", also called "predictive power" (see Whitehead et al (2008)) and "assurance" (see O'Hagan et al (2005)). The basic idea is to pre-specify a value for the average power and use this as the basis for sample sizing. (This pre-specified value must be chosen with care; it does not seem to be widely appreciated that, as the sample size increases, the average power is bounded above by the prior probability that the new treatment is better than the control.) A frequentist analysis will then follow, after the trial data has been collected. In the decision-theoretic Bayesian approach, "sample size is based on maximization of a utility function, formulated to reflect concerns involving mainly measures of cost and benefit" (Whitehead et al (2008)). Because of the difficulties involved in specifying utility functions in practice and because "misspecification of utilities can lead to seriously sub-optimal designs", Whitehead et al (2008) conclude that "implementation (of decision-theoretic Bayesian approaches) in clinical trial design has been and probably will remain limited". We agree with this assessment.

In a "proper Bayesian" approach to sample size determination, it is assumed that a Bayesian analysis will be performed at the end of the trial, and that the analysis will generally be based on whatever criterion (in terms of a posterior distribution) has been specified at the sample size determination stage. Whether or not this is actually done however (and, as is pointed out in O'Hagan et al (2005), regulatory agencies have been slow to adopt formal guidelines encouraging Bayesian analyses in Phase III drug confirmatory trials), the use of Bayesian concepts at the design stage leads to increased understanding of how sample size is affected by the specification of prior distributions which incorporate prior knowledge. An approach proposed by Whitehead et al (2008) for exploratory clinical trials falls into the "proper Bayesian" category, as do methods based on credible intervals, such as average coverage and average length (see e.g. Joseph and Bélisle (1997) and M'Lan et al (2008), and the references therein). It should also be pointed out that proposals have been made to utilize two different priors for the different stages of design and analysis (see e.g. Brutti et al (2008) and O'Hagan and Stevens (2001)). In the latter paper, the "analysis prior" is essentially non-informative, so that the analysis more closely resembles a standard frequentist one. We do not pursue this approach here, although it is straightforward to implement.

This paper is structured as follows. In Section 2 we describe a "proper Bayesian" approach to sample size determination that seems intuitively appealing; it is based on choosing a sample size which gives a pre-specified "probability of a successful trial", where the probability is calculated using the marginal distribution of the data. Section 3 is concerned with specific examples and applications in a common clinical trial setting, that of comparing the means of two normally distributed samples.

## 2. The "Probability of a Successful Trial" Criterion

Assume that, in a clinical trial, we will collect data $X_1, \ldots, X_n$ on a total of $n$ subjects, where $X^{(n)} \equiv (X_1, \ldots, X_n)$ has a joint distribution that depends on a parameter $\theta \in \Theta$, where $\Theta$ is a subset of a multidimensional Euclidean space. The variable $X_i$ being measured or calculated for final analysis is often referred to as the "primary endpoint" of the trial. This is usually 1-dimensional, but may be multivariate if there are two or more primary endpoints. Let $\Theta_0$ and $\Theta_1$ be two disjoint, non-empty subsets of $\Theta$, with $\Theta_0 \cup \Theta_1 = \Theta$. (In a classical hypothesis testing framework, these subsets correspond to null and alternative hypotheses.) We are interested in concluding that $\theta \in \Theta_1$.

Let $\pi$ denote a prior distribution for $\theta$, and denote by $\pi_n(\cdot \mid x^{(n)})$ the posterior distribution of $\theta$ given the observed data $x^{(n)} \equiv (x_1, \ldots, x_n)$. In the notation being used here, distributions refer to probability measures. (In the examples in Section 3, these distributions will have probability density functions.) Thus, for example, $\pi(\Theta_1)$ and $\pi_n(\Theta_1 \mid x^{(n)})$ are respectively the prior and posterior probabilities that $\theta \in \Theta_1$.

Suppose that, at the analysis stage (that is, after the data $x^{(n)}$ has been observed), we call a clinical trial a *success* if the posterior probability that $\theta \in \Theta_1$ is greater than or equal to a specified threshold $\eta \in (0,1)$; that is, if

$$\pi_n\left(\Theta_1 \mid x^{(n)}\right) \geq \eta. \tag{1}$$

At the design stage, since we haven't yet collected the trial data, we cannot evaluate the left side of (1). We can, however, assess the inequality (1) using the marginal (or unconditional) distribution, $P_{X^{(n)}}$ (say), of $X^{(n)}$, obtained from the joint distribution of $X^{(n)}$ (given $\theta$) and the prior $\pi$ by integrating out $\theta$. (In the Bayesian sample size determination literature, this distribution has also been called the "predictive distribution"; see, e.g. Joseph and Bélisle (1977) and Brutti et al (2008).) We can ask: What is the probability, according to the marginal distribution $P_{X^{(n)}}$, that the trial will be a success? Formally, this involves calculating the quantity

$$\psi(n) = P_{X^{(n)}}(X^{(n)} \in E_n), \tag{2}$$

where $E_n$ denotes the set of all samples $x^{(n)}$ which lead to a successful trial, namely

$$E_n = \left\{ x^{(n)}; \pi_n(\Theta_1 \mid x^{(n)}) \geq \eta \right\}. \tag{3}$$

We will refer to $\psi(n)$ in (2) as the *probability of a successful trial* and abbreviate it as PST. (It has also been called "expected Bayesian power" by Spielgelhalter et al (2004), and "predictive probability" by Brutti et al (2008).) The basic idea now is to choose a sample size $n$ for which the PST $\psi(n)$ is "large enough", in the sense that it exceeds a specified threshold. As we note below, there is a limitation on the size of this threshold. It is clear that, for fixed $n$, $\psi(n)$ increases as $\eta$ in (1) decreases; that is, the lower the threshold for defining a successful trial, the greater the chance of obtaining a sample leading to one.

Although we have emphasized the dependence in (2) of $\psi(n)$ on $n$, it of course also depends on any hyperparameters in the prior $\pi$, as well as the subset $\Theta_1$ deemed appropriate and the choice of the threshold $\eta$. In practical situations, we would choose $\eta$ to be "large", with the choice influenced by the specification of $\Theta_1$, and by the type of trial. In the examples in Section 3, $\Theta_1$ will correspond to a classical alternative hypothesis, and we will take $\eta = 1 - \alpha$, where $\alpha$ is a standard significance level. (The choice of significance level is influenced by the type of trial – it is generally taken to be smaller for a confirmatory (Phase III) trial than for an exploratory or earlier phase trial.) As previously noted, the PST increases as $\eta$ decreases, and this will be illustrated in Section 3.

How large can the PST in (2) be? This raises the non-trivial question of how $\psi(n)$ behaves as for large $n$. In the examples in Section 3, for the parameter values considered, $\psi(n)$ is an increasing function of *n*, for all *n*. In the first example in Section 3 (two normal samples with unknown means and known common variance), it is straightforward to show that

$$\lim_{n \to \infty} \psi(n) = \pi(\Theta_1); \tag{4}$$

that is, no matter what value of $\eta \in (0,1)$ is chosen in (1) to define a successful trial, as $n \to \infty$, the PST $\psi(n)$ approaches the prior probability that $\theta \in \Theta_1$. The limiting result (4) has been obtained by Brutti et al (2008) in the case of a single normal sample with unknown mean and known variance. It is established in Eaton et al (2012) that (4) holds much more generally, with the proof hinging on consistency of the posterior distribution. (For a discussion of Bayesian consistency, see Ghosh and Ramamoorthi (2003)). Thus, even with an infinite sample size (where we would learn the "true" value of $\theta$), the PST cannot exceed the prior probability that $\theta \in \Theta_1$. Because of this, it may be more informative, when choosing a sample size, to focus attention on the "normalized PST index"

$$\psi^*(n) = \frac{\psi(n)}{\pi(\Theta_1)}, \tag{5}$$

for which $\lim_{n \to \infty} \psi^*(n) = 1$, so that $\psi^*(n)$ represents the proportion of the maximum value of the PST explained by the sample size *n*. We do this in the examples in the next section.

### 3. Examples

Throughout this section, we consider a normal model for comparing the means of two populations, corresponding to experimental treatment *E* and control *C*. We base our notation loosely on that used by Whitehead et al (2008). We have independent samples of sizes $n_E$ and $n_C$ from each population, so that the model (conditional on parameters) is

$$X_1, \ldots, X_{n_E} \sim iid \, N\left(\mu_E, \tau^{-1}\right), \quad Y_1, \ldots, Y_{n_C} \sim iid \, N\left(\mu_E, \tau^{-1}\right). \tag{6}$$

We assume that the precision $\tau$ is the same in both populations. Let $n = n_E + n_C$ denote the total sample size, and let $R = n_E / n_C$ be the allocation ratio. In terms of *n* and *R*, the two sample sizes are $n_E = nR / (1+R), n_C = n / (1+R)$. Typically we fix a value of *R*, and then the two group sample sizes follow from the choice of *n*. (In the illustrations below, we will always take *R*=1, so that there are equal sample sizes in the two groups.) The parameter of interest here is the *treatment effect* $\delta = \mu_E - \mu_C$, and positive values of $\delta$ are assumed to favor the experimental treatment *E* over the control *C*. We base our definition of a "successful trial" (see (1)) on the posterior probability that $\delta > 0$. (It is straightforward to modify this to $\delta > \delta^*$ (for example) if this seems more appropriate in a particular situation.)

**3.1 Known precision, conjugate prior:** When $\tau$ is known, a standard conjugate prior specifies that the two means $\mu_E$ and $\mu_C$ are independent, with

$$\mu_E \sim N\left(\mu_E^{(0)}, \frac{1}{n_E^{(0)}\tau}\right), \quad \mu_C \sim N\left(\mu_C^{(0)}, \frac{1}{n_C^{(0)}\tau}\right). \tag{7}$$

Here $\mu_E^{(0)}, \mu_C^{(0)}, n_E^{(0)}, n_C^{(0)}$ are specified, and the superscript (0) is used to indicate a quantities in (or calculated from quantities in) prior distributions. (In what follows, the superscript (1) will similarly indicate a quantity in a posterior distribution.) The $n^{(0)}$'s multiplying the precision play the role of "pseudo observations", and allow us to weight prior information differently in the groups $E$ and $C$. We might, for example have much more information about the control group $C$ from previous trials or historical data, in which case we might choose $n_C^{(0)}$ to be large in comparison with $n_E^{(0)}$. The aim in such a situation would be to allocate more subjects to group $E$ (allocation ratio $R > 1$).

With the Bayesian model specified by (6) and (7), the posterior distribution of $\delta = \mu_E - \mu_C$ depends only on the trial results through the observed sample means $\bar{x}$ and $\bar{y}$, and is

$$\delta \mid \bar{x}, \bar{y} \sim N\left(\delta^{(1)}, \frac{1}{D^{(1)}\tau}\right), \tag{8}$$

where the posterior mean $\delta^{(1)}$ is the difference between the posterior means of $\mu_E$ and $\mu_C$, namely

$$\delta^{(1)} \equiv \delta^{(1)}\left(\bar{x}, \bar{y}\right) = \mu_E^{(1)} - \mu_C^{(1)}, \tag{9}$$

with

$$\mu_E^{(1)} = \frac{1}{n_E^{(1)}}\left(n_E^{(0)}\mu_E^{(0)} + n_E\bar{x}\right), \quad n_E^{(1)} = n_E^{(0)} + n_E,$$
$$\mu_C^{(1)} = \frac{1}{n_C^{(1)}}\left(n_C^{(0)}\mu_C^{(0)} + n_C\bar{y}\right), \quad n_C^{(1)} = n_C^{(0)} + n_C, \tag{10}$$

and $D^{(1)}$ in (8) is

$$D^{(1)} = \frac{n_E^{(1)}n_C^{(1)}}{n_E^{(1)} + n_C^{(1)}}. \tag{11}$$

From (1), we call the trial a success if the posterior probability that $\delta > 0$ exceeds a specified threshold $\eta$. This condition is equivalent to

$$\delta^{(1)}\sqrt{D^{(1)}\tau} \geq z_\eta,$$ (12)

where $z_\eta$ denotes the $\eta$-quantile of the $N(0,1)$ distribution. Then the PST is (see (2))

$$\psi(n) = P\left\{\delta^{(1)} \geq \frac{z_\eta}{\sqrt{D^{(1)}\tau}}\right\},$$ (13)

where the probability in (13) is calculated using the marginal distribution of $\delta^{(1)}(\bar{X},\bar{Y})$. This distribution is $N(\Delta,\sigma^2)$, with mean and variance given by

$$\Delta = \mu_E^{(0)} - \mu_C^{(0)}, \quad \sigma^2 = \frac{n_E}{\tau n_E^{(0)} n_E^{(1)}} + \frac{n_C}{\tau n_C^{(0)} n_C^{(1)}}.$$ (14)

Consequently,

$$\psi(n) = \Phi\left(\frac{1}{\sigma}\left(\Delta - \frac{z_\eta}{\sqrt{D^{(1)}\tau}}\right)\right).$$ (15)

(Note that this depends on the prior means only through their difference $\Delta$.) Since the prior probability that $\delta > 0$ is

$$P(\delta > 0) = \Phi\left(\Delta\sqrt{\tau D^{(0)}}\right),$$ (16)

where

$$D^{(0)} = \frac{n_E^{(0)} n_C^{(0)}}{n_E^{(0)} + n_C^{(0)}},$$ (17)

the normalized PST index (see(5)) is

$$\psi^*(n) = \frac{\Phi\left(\frac{1}{\sigma}\left(\Delta - \frac{z_\eta}{\sqrt{D^{(1)}\tau}}\right)\right)}{\Phi\left(\Delta\sqrt{\tau D^{(0)}}\right)}.$$ (18)

*Example 3.1.1:* The "International Restless Legs Syndrome Study Group Rating Scale" (IRLS) is used in clinical trials that attempt to show that a drug is efficacious in treating "Restless Legs Syndrome" (RLS). The IRLS is a clinician-administered 10-item questionnaire used to assess the severity of RLS. The overall score ranges from 0 to 40, with lower scores reflecting lower severity and better quality of life. We consider a randomized trial aimed at comparing a drug group with a placebo group, and take as the endpoint the difference between the IRLS score at baseline and at the end of the trial. If the drug is efficacious, we would expect to see positive values of this difference in the drug group.

For the purpose of illustration here, assume the standard deviation (SD) of the endpoint in both populations is known to be 8. (This is about equal to the average SD observed in Allen et al (2010), a published RLS study.) In a frequentist approach, assume a difference in mean scores (drug vs. placebo) of 4 is "clinically meaningful". In a confirmatory Phase III trial, a test of "no difference between drug and placebo" is usually a 2-sided, 0.05-level test (even when the hope is to conclude that drug is better than placebo); consequently we will focus here on a 1-sided 0.025-level test. A standard normal-based calculation shows that, in order to have 80% power at $\delta^* = 4$, we need 64 subjects in each group, for a total sample size of 128.

Figure 1 illustrates how the normalized PST index (18) behaves as a function of total sample size $n$ for four different values of the difference in prior means $\Delta$, when $R = 1$, $\eta = 0.975$ and the prior weights are taken to be $n_E^{(0)} = n_C^{(0)} = 2$ (in which case the prior distribution of the treatment effect $\delta$ is $N(\Delta, 64)$). As noted in Section 2, the curves all converge to 1 as the sample size increases, although this is not obvious from Figure 1, which only goes to $n = 200$. In a practical setting, we would probably choose $\Delta$ to reflect our prior opinion, perhaps based on previous studies, about the treatment effect. Alternatively, we might take it to be the "clinically meaningful" treatment effect. If we take, for example, $\Delta = 4$, the PST converges to a maximum of $\Phi(0.5) = 0.6915$. The normalized PST index gives the proportion of this maximum value achieved for various sample sizes – see Table 1. We see, for example, that when $n = 100$ (so there are 50 subjects in each group, the PST is 79% of its maximum value, and it is increasing only slowly with $n$. There is not much to be gained, in terms of the PST, by taking larger sample sizes. Also, we noted in Section 2 that, for fixed $\Delta$, the PST increases as the threshold $\eta$ decreases (and then, of course, a smaller sample size may be chosen) – see Figure 2.

When the prior weights $n_E^{(0)}$ and $n_C^{(0)}$ are increased (so that the prior distributions for the means are sharper or more informative) we expect the PST to also increase. The second set of two rows in Table 1, with $n_E^{(0)} = n_C^{(0)} = 30$, illustrates this. The PST values are considerably higher here than in the first set; the normalized PST index is 0.80 with a total sample size of 60.

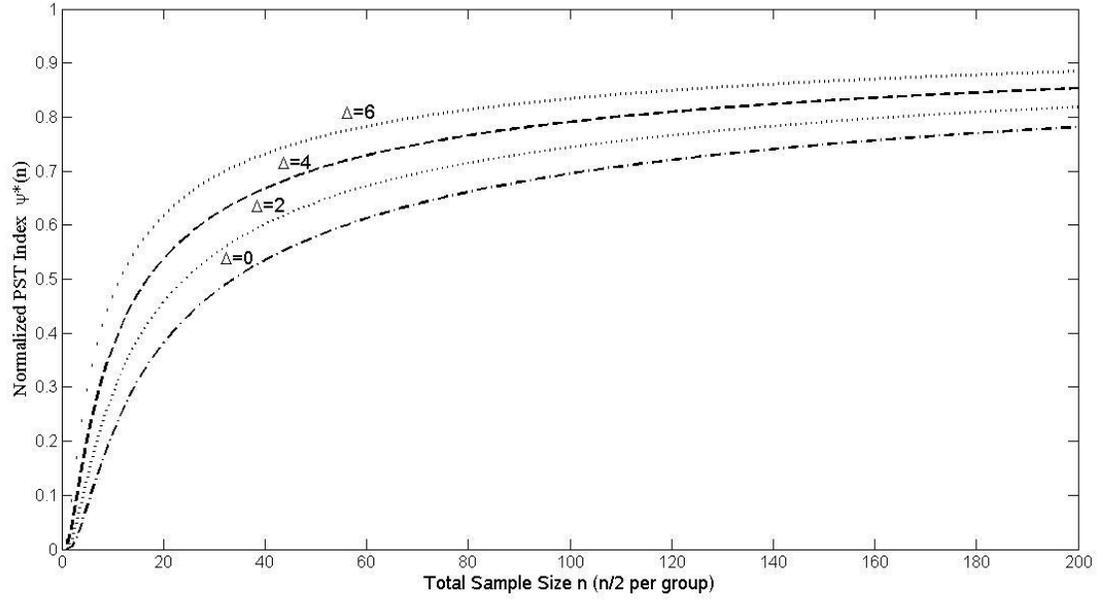

Figure 1: Normalized PST index versus $n$ for various values of $\Delta$. ( $\eta = 0.975$  SD = 8, $n_E^{(0)} = n_C^{(0)} = 2$ )

| | $n$ | 40 | 60 | 80 | 100 | 120 | 140 |
|---|---|---|---|---|---|---|---|
| $n_E^{(0)} = n_C^{(0)} = 2$ | $\psi(n)$ | 0.46 | 0.50 | 0.53 | 0.55 | 0.56 | 0.57 |
| | $\psi^*(n)$ | 0.67 | 0.73 | 0.77 | 0.79 | 0.81 | 0.83 |
| $n_E^{(0)} = n_C^{(0)} = 30$ | $\psi(n)$ | 0.75 | 0.78 | 0.81 | 0.82 | 0.84 | 0.85 |
| | $\psi^*(n)$ | 0.77 | 0.80 | 0.83 | 0.85 | 0.86 | 0.87 |

Table 1: Values of the PST and the normalized PST index for two sets of values of $n_E^{(0)}$ and $n_C^{(0)}$ when $\Delta = 4$. ( $\eta = 0.975$, SD = 8)

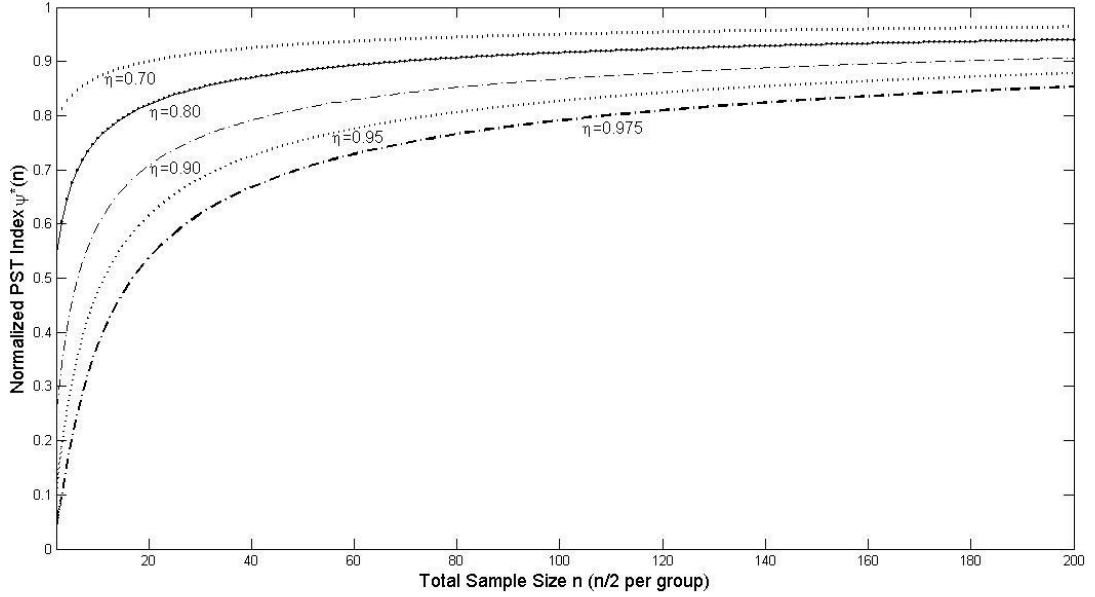

Figure 2: Normalized PST index versus $n$ for various values of the threshold $\eta$ ($\Delta = 4$, SD $= 8$, $n_E^{(0)} = n_C^{(0)} = 2$)

**3.2 Known precision, mixture prior:** Suppose that we are in a situation where we believe that there is some chance (perhaps small) that the treatment effect $\delta$ is "close" to zero. This might arise, for example, when a prior is based on the elicitation of opinions of experts, some of whom are skeptical about the efficacy of a proposed new drug, whereas others are more confident. In the frequentist setting, when testing the null hypothesis $\delta = 0$, sufficiency and translation invariance leads us to consider $U \equiv \bar{X} - \bar{Y} \sim N\left(\delta, \left(\tau_n^*\right)^{-1}\right)$, with $\tau_n^* = n_E n_C \tau / n$. For the purpose of illustration, assume that the prior distribution for $\delta$ is a mixture of two normal distributions, $N\left(0, \tau_0^{-1}\right)$ and $N\left(\delta_1, \tau_1^{-1}\right)$, with mixing probabilities $\rho$ and $1 - \rho$. Then, given $U = u$, the posterior probability that $\delta > 0$ is

$$P\left(\delta > 0 \mid u\right) = \tilde{\rho}(u) \Phi\left(\frac{\tau_n^* u}{\sqrt{\tau_n^* + \tau_0}}\right) + \left(1 - \tilde{\rho}(u)\right) \Phi\left(\frac{\tau_n^* u + \tau_1 \mu_1}{\sqrt{\tau_n^* + \tau_0}}\right), \tag{19}$$

where

$$\tilde{\rho}(u) = \frac{\rho}{f(u)} \times \left\{ N\left(u; 0, \frac{\tau_n^* + \tau_0}{\tau_n^* \tau_0}\right) \right\}, \tag{20}$$

and where

$$f(u) = \rho \times \left\{ N\left(u; 0, \frac{\tau_n^* + \tau_0}{\tau_n^* \tau_0}\right) \right\} + (1-\rho) \times \left\{ N\left(u; \mu_1, \frac{\tau_n^* + \tau_1}{\tau_n^* \tau_1}\right) \right\} \tag{21}$$

is the density function of the marginal distribution of $U$. (In (20) and (21), the notation $N\left(u; \mu, \sigma^2\right)$ denotes the $N\left(\mu, \sigma^2\right)$ density function evaluated at $u$.) Then the PST is $\psi(n) = P(U \in E_n)$, where $E_n$ is the set of all values $u$ of $U$ for which the posterior probability (19) exceeds a specified value $\eta$. Although an analytic expression for $\psi(n)$ seems elusive, it is a straightforward matter to calculate it via Monte Carlo simulation, and a MATLAB program which does this is available from the authors.

*Example 3.2.1*: This is a continuation of Example 3.1.1. When $\delta$ has the mixture prior described above, the prior mean and variance are

$$E(\delta) = (1-\rho)\delta_1, \quad \mathrm{Var}(\delta) = \rho(1-\rho)\delta_1^2 + \frac{\rho}{\tau_0} + \frac{1-\rho}{\tau_1}. \tag{22}$$

For the purpose of illustration here, we take $\tau_0 = 100$, $E(\delta) = 4$, and, as in Example 3.1.1, $\mathrm{Var}(\delta) = 64$. Given $\rho$, we can then determine the parameters $\delta_1$ and $\tau_1$ in the second mixture distribution appropriately. Figure 3 gives the graphs for the normalized PST index for various values of $\rho$. If, for example, we take $\rho = 0.1$ (in which case $\delta_1 = 4.44$, $\tau_1^{-1} = 69.13$), the limiting value of the PST is 0.6830. The normalized PST index gives the proportion of this maximum value achieved for various sample sizes – see Table 2.

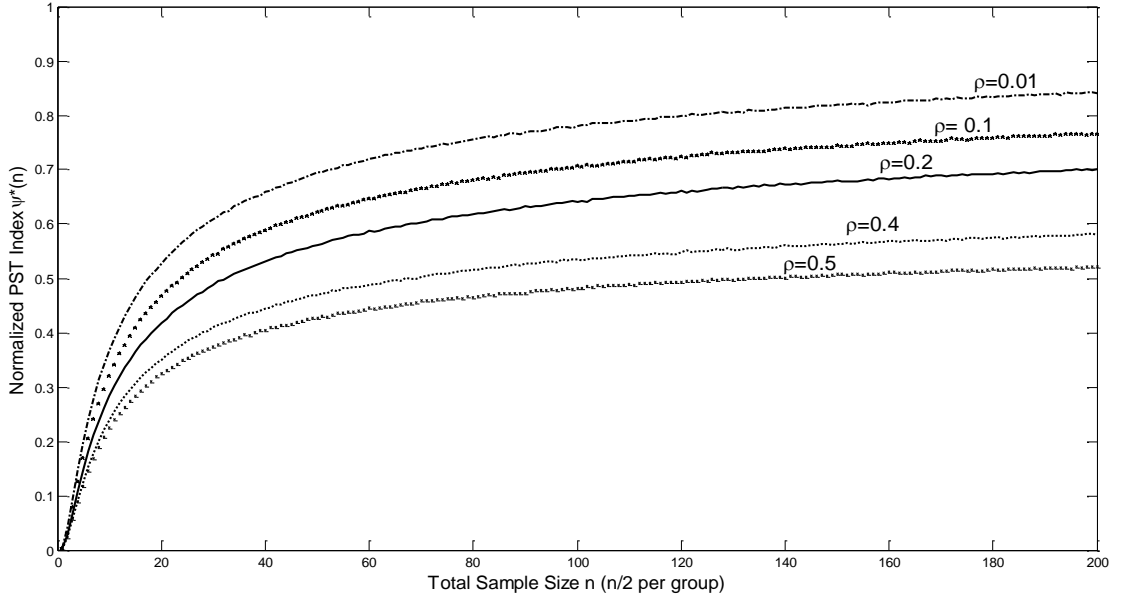

Figure 3: Normalized PST index versus $n$ for various values of $\rho$, and $\eta = 0.975$. (Each mixture has mean 4 and SD 8.)

| $n$ | 20 | 40 | 60 | 80 | 100 | 120 | 140 |
|---|---|---|---|---|---|---|---|
| $\psi(n)$ | 0.32 | 0.40 | 0.44 | 0.46 | 0.48 | 0.49 | 0.50 |
| $\psi^*(n)$ | 0.47 | 0.59 | 0.65 | 0.68 | 0.71 | 0.72 | 0.74 |

Table 2: Values of the PST and the normalized PST index when $\rho = 0.1$.

**3.3 Unknown precision, conjugate prior:** We begin with the model in Section 3.1, but now assume that the precision $\tau$ is unknown, and that its prior distribution is gamma, with specified hyperparameters $\alpha^{(0)}$ and $\beta^{(0)}$ (so that $E(\tau) = \alpha^{(0)}/\beta^{(0)}$ and $\mathrm{Var}(\tau) = \alpha^{(0)}/\beta^{(0)2})$ . Given $\tau$, the means $\mu_E$ and $\mu_C$ are independent, with conditional prior distributions given by (7). Then the posterior distribution of $\tau$ (see, e.g. Whitehead et al (2008)) is gamma, with parameters $\alpha^{(1)}$ and $\beta^{(1)}$, where

$$\alpha^{(1)} = \alpha^{(0)} + \tfrac{1}{2}n, \quad \beta^{(1)} \equiv \beta^{(1)}\left(\overline{x}, \overline{y}, s^2\right) = \beta^{(0)} + \tfrac{1}{2}H, \tag{23}$$

with

$$H \equiv H\left(\bar{x}, \bar{y}, s^2\right) = (n-2)s^2 + \frac{n_E n_E^{(0)}}{n_E^{(1)}}\left(\bar{x} - \mu_E^{(0)}\right)^2 + \frac{n_C n_C^{(0)}}{n_C^{(1)}}\left(\bar{y} - \mu_C^{(0)}\right)^2 \qquad (24)$$

and $s^2$ being the value of the (unbiased) pooled sample variance $S^2$. Next, define $T$ as

$$T = \sqrt{\frac{D^{(1)} \alpha^{(1)}}{\beta^{(1)}}}\left(\delta - \delta^{(1)}\right), \qquad (25)$$

where $\delta^{(1)}, D^{(1)}, \alpha^{(1)}$ and $\beta^{(1)}$ are given in (9), (11), and (23). Then (see Joseph and Bélisle (1997), Whitehead et al (2008)) the posterior distribution of $T$ is $t_{2\alpha^{(1)}}$. We call the trial a success if

$P\left(\delta > 0 \mid \bar{x}, \bar{y}, s^2\right) \geq \eta$. Using (25) this is equivalent to

$$\delta^{(1)} \sqrt{\frac{D^{(1)} \alpha^{(1)}}{\beta^{(1)}}} \geq t_{2\alpha^{(1)}, \eta}, \qquad (26)$$

where the right side of (26) denotes the $\eta-$quantile of the $t_{2\alpha^{(1)}}$ distribution. Then the PST is

$$\psi(n) = P\left\{\delta^{(1)}\left(\bar{X}, \bar{Y}\right) \sqrt{\frac{D^{(1)} \alpha^{(1)}}{\beta^{(1)}\left(\bar{X}, \bar{Y}, S\right)}} \geq t_{2\alpha^{(1)}, \eta}\right\} \qquad (27)$$

where the probability in (27) is calculated using the marginal joint distribution of $\bar{X}, \bar{Y}$, and $S^2$. It may be shown that (27) depends on the prior means only through their difference $\Delta$. Although it appears difficult to obtain an analytical expression, $\psi(n)$ in (27) can be calculated via simulation, by first conditioning on the model parameters. A MATLAB program that does this is available from the authors.

*Example 3.3.1:* This continues Example 3.1.1. We use six precision estimates for IRLS change from baseline (mean 0.015 and SD of 0.0010) from Table 2 of Allen et al (2010) to fit a prior gamma distribution with $\alpha^{(0)} = 243$ and $\beta^{(0)} = 16200$. Figure 4 gives the normalized PST curves for various values of $\Delta$ when $\eta = 0.975$ and $n_E^{(0)} = n_C^{(0)} = 2$ (as in Figure 1). The mean precision here corresponds to a SD of about 8.16 (Example 3.1.1 assumed a known SD of 8), and the precision variability is small, so that the graphs in Figure 4 are very similar to those in Figure 1.

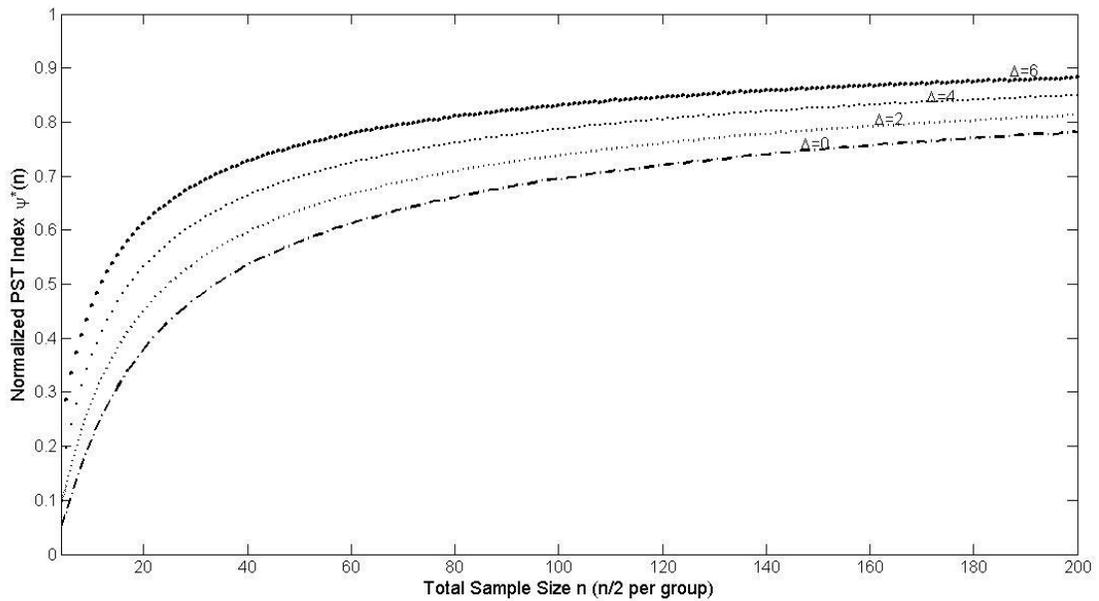

Figure 3. Normalized PST index versus *n* for various values of $\Delta$.

## 4. Summary

   We have described a "proper Bayesian" approach to sample size determination, and illustrated its use in a context of a two parallel arm clinical trial for superiority with normal data. It is more widely applicable --- for example, in non-inferiority and equivalence trials, and to non-normal data (such as dichotomous data and time-to-event-data). In most instances, the proposed PST criterion will have to be calculated via simulation, and – at least in standard distributional settings – this appears to be reasonably straightforward. The material in Section 3 may also be readily generalized to the multiple primary endpoints setting, although the number of hyperparameters that have to be specified in prior conjugate distributions grows rapidly as the dimension increases.

## Acknowledgments

The authors appreciate helpful comments provided by a referee. We also thank Phil Woodward for insightful remarks on an earlier draft of this paper.